\documentclass[aps,prb,amssymb,twocolumn,groupedaddress,floats]{revtex4}

\usepackage{epsfig}
\usepackage{amsmath}

\begin{document}
\draft
\author{D.\ Meyer}
\email[]{Dietrich.Meyer@physik.hu-berlin.de}
\author{W.\ Nolting}
\affiliation{Lehrstuhl Festk{\"o}rpertheorie, Institut f{\"u}r Physik,
  Humboldt-Universit{\"a}t zu Berlin, Invalidenstr.\ 110, 10115 Berlin}



\title{Kondo screening and exhaustion in the periodic Anderson model}

\begin{abstract}
We investigate the paramagnetic periodic Anderson model using the
dynamical mean-field theory in combination with the modified
perturbation theory which interpolates between the weak and strong
coupling limits. For the symmetric PAM, the ground state is always a
singlet state. However, as function of the hybridization strength, a
crossover from collective to local Kondo screening is found. Reducing
the number of conduction electrons, the local Kondo singlets remain
stable. The unpaired $f$-electrons dominate the physics of the system.
For very low conduction electron densities, a large increase of the
effective mass of the quasiparticles is visible, which is interpreted as
the approach of the Mott-Hubbard transition.
\end{abstract}
\maketitle

\section{Introduction} 
The periodic Anderson model (PAM) is one of the standard models for
heavy fermion systems\cite{hewson}. In its simplest form, it describes a
system of localized electronic states which hybridize with an
uncorrelated conduction band. Except for a few statements concerning
ground state properties\cite{TSU97}, no exact solution of the PAM is
known up to now. There have been many approximate calculations like
QMC\cite{Jar95,TJF97,TJF98,GE98,GE99,GE99pre2,HMS99}, perturbation
theory\cite{LM78,SC90a,SC90b} as well as other analytical
approaches
\cite{DS97,CS97,DS98,MW93,VGR92,VGR94,VTJK99pre,PBJ99pre,MNRR98,MN99a,MN99pre}.
However, it is still far from being understood. Closely related to the
PAM is the single-impurity Anderson model (SIAM)\cite{hewson}.
The SIAM is one of the best-understood models in theoretical physics;
besides a broad range of approximate solutions even some exact
calculations are possible (for a detailed review, see reference
\onlinecite{hewson}). The main result of these calculations is the
emergence of a new temperature scale $T_K$ (\textit{Kondo Temperature}),
which governs the low-temperature physics.
For $T<T_K$, the magnetic
moment of the impurity is screened by conduction electrons
(\textit{Kondo screening}). All thermodynamic properties can be
expressed in terms of $T_K$.

However, due to the periodicity of the localized states in the case of
the PAM, new and more complicated physical properties will emerge.  The
most prominent example is the so-called exhaustion problem first
mentioned by Nozi\`{e}res\cite{Noz98}.  His argumentation was based on
the Kondo lattice model (KLM), which can be derived from the PAM under
the condition of half-filling for the localized states using the
Schrieffer-Wolff transformation\cite{SW66,hewson}. In the KLM, where the
charge degrees of freedom of the localized states have been removed,
Kondo screening manifests itself in the quenching of the magnetic moment
of the localized spins by the conduction electrons.  Reducing the number
of conduction electrons below half-filling, however, the situation
changes. The conduction band occupation is not sufficient to screen the
localized spins completely.  The nature of the ground state is not
clear.  As noted by Nozi\`{e}res\cite{Noz98}, the situation is different
for the case of small and large $J$, respectively.  Whereas for small
$J$, the screening is a collective effect, in the limit
$J\rightarrow\infty$, it can be understood as the formation of local
Kondo singlets, in which one conduction electron and one
$S=\frac{1}{2}$-spin form a bound singlet state.  On removing conduction
electrons, some of these local Kondo singlets (LKS) will be broken.  The
remaining ``bachelor'' spins can be described as a system of hard-core
fermions, for which double occupancy is forbidden\cite{Noz98}. The
small-$J$ limit with its collective screening is more complicated. But
nevertheless, following Nozi\`{e}res, this case can similarly be mapped
onto an effective Hubbard model\cite{Noz98}.

Contrary to the above-described KLM, the localized states also have a
charge degree of freedom in the PAM, and its physical properties are
therefore more complicated. The Schrieffer-Wolff transformation is valid
only in the small $J=\frac{V^2}{U}$- limit ($V$ is the hybridization
strength between localized and the conduction band states and $U$ is the
on-site Coulomb interaction among the localized electrons (see section
\ref{sec:theo}). So, especially in the limit of strong hybridization
between localized and conduction states, the KLM is not \textit{a
priori} justified as an effective model for the PAM. For the weakly
hybridized case, the exhaustion in the PAM has recently been discussed
by several authors\cite{TJF97,TJF98,VTJK99pre,PBJ99pre}. They find that
Nozi\`{e}res' picture of the effective Hubbard model also holds in this
case and explains the emergence of a new low-temperature scale.

In this paper, we want to present a systematic investigation of the
PAM. In the following chapter,we will introduce the theoretical
approach, which combines the dynamical mean-field theory
(DMFT)\cite{MV89,PJF95,GKKR96} with a
modification of the iterated perturbation theory
(IPT)\cite{KK96,PWN97,WPN98,MWPN99}. The results will
be presented in section~\ref{sec:results-discussion}. In the first part
of the discussion, we will focus on the hybridization-strength
dependence of the symmetric PAM, and in the second part, we reduce the
number of conduction electrons and investigate the exhaustion problem,
thereby extending the discussion of reference~\onlinecite{VTJK99pre}.

\section{Theory}
\label{sec:theo}
The PAM is defined by its Hamiltonian
\begin{align} 
  \label{hamiltonian} 
    H =&\sum_{\vec{k},\sigma} 
\epsilon(\vec{k})s_{\vec{k},\sigma}^{\dagger}s_{\vec{k},\sigma} + 
    \sum_{i,\sigma} e_f f_{i,\sigma}^{\dagger}f_{i,\sigma} +\\ &V 
    \sum_{i,\sigma} (f_{i,\sigma}^{\dagger}s_{i,\sigma} + 
    s_{i,\sigma}^{\dagger}f_{i,\sigma} ) + \frac{1}{2} U \sum_{i,\sigma} 
    n_{i,\sigma}^{(f)}n_{i,-\sigma}^{(f)}\nonumber 
\end{align} 
with $s_{i,\sigma}$ ($f_{i,\sigma}$) and $s_{i,\sigma}^{\dagger}$
($f_{i,\sigma}^{\dagger}$) being the conduction band ($f$-level)
electron annihilation and creation operators
($n_{i,\sigma}^{(f)}=f_{i,\sigma}^{\dagger}f_{i,\sigma}$).
$\epsilon(\vec{k})$ is the dispersion of a non-degenerate $s$-type
conduction band and $e_f$ denotes the position of the $f$-level with
respect to the center of gravity of the conduction band. The
hybridization $V$ is taken as a $\vec{k}$-independent constant, and
finally $U$ is the local Coulomb interaction between two $f$-electrons
at the same lattice site.

To obtain the the $f$-electron Green function $G_{ii\sigma}^{(f)}(E)=
\langle\!\langle f_{i\sigma};f_{i\sigma}^{\dagger}\rangle\!\rangle$, we
apply a two-step procedure.  The first of these, known as
\textit{dynamical mean-field theory} (DMFT)\cite{MV89,PJF95,GKKR96}, is
a mapping of the PAM onto a simpler model, namely the single-impurity
Anderson model. The second step of our procedure is to find an
approximate solution of the SIAM using the modified perturbation
theory (MPT)\cite{PWN97,WPN98,MWPN99}.

The starting point of the DMFT is the assumption of a
$\vec{k}$-independent, i.e. local self-energy $\Sigma_{\sigma}(E)$. 
It can be shown that in this case, the self-energy of the PAM is
equivalent to the self-energy of a properly defined impurity model
(SIAM). The conduction band within this SIAM has to be determined by the 
following expression for the hybridization function\cite{hewson}
\begin{equation}
  \label{eq:seco}
  \Delta_{\sigma}(E)= E-e_f-\Sigma_{\sigma}(E)
  -\left(G_{ii\sigma}^{(f)}(E)\right)^{-1} 
\end{equation}
In the original SIAM, this function is given as
$\Delta(E)=\sum_{\vec{k}} \frac{V^2}{E-\epsilon(\vec{k})}$. All
information of the conduction band and its hybridization with the
impurity is contained in $\Delta(E)$. Therefore its knowledge is
sufficient to define the electron bath of the SIAM also within the
DMFT.

From the self-energy $\Sigma_{\sigma}(E)$ of the SIAM defined by
$\Delta_{\sigma}(E)$, the PAM $f$-electron Green function can directly be
obtained as
\begin{equation}
  \label{eq:gf}
  G_{ii\sigma}^{(f)}(E)= \frac{1}{N} \sum_{\vec{k}}
  \frac{1}{E-e_f-\frac{V^2}{E-\epsilon(\vec{k})}-\Sigma_{\sigma}(E)}    
\end{equation}
Since $G_{ii\sigma}^{(f)}(E)$ enters expression (\ref{eq:seco}), a
self-consistent solution has to be found by iteration. As already noted
above, the DMFT-procedure becomes exact for a local, i.\ e.\
$\vec{k}$-independent self-energy. It has been shown that for the limit
of infinite dimensions, or equivalently in the limit of the lattice
coordination number going to infinity, this is indeed exactly the
case\cite{MV89,Mue89}. Furthermore, solving the PAM in three dimensions
using 
perturbation theory, it was shown that the results obtained in the local 
approximation compare qualitatively and quantitatively very well with
those where the full $\vec{k}$-dependence has been
considered\cite{SC89b,SC90a}.

Now the actual problem is shifted to solve the SIAM which is defined by
$\Delta_{\sigma}(E)$ (see equation (\ref{eq:seco})). 
For this task we apply the modified perturbation theory
(MPT)\cite{PWN97,WPN98,MWPN99} which
is based on the following 
ansatz for the electronic self-energy\cite{MR82,KK96}
\begin{equation}
  \label{eq:ansatz}
  \Sigma_{\sigma}(E)=U \langle n_{-\sigma}^{(f)}\rangle
  +\frac{a_{\sigma} \Sigma_{\sigma}^{\rm (SOC)}(E)}
  {1-b_{\sigma} \Sigma_{\sigma}^{\rm (SOC)}(E)}
\end{equation}
where $\Sigma_{\sigma}^{\rm (SOC)}(E)$ denotes the second-order
contribution to the second-order perturbation theory around the
Hartree-Fock solution (SOPT-HF)\cite{Yam75,ZH83,SC90a}.
Please note that the ansatz~(\ref{eq:ansatz}) is
$\vec{k}$-independent by
construction, the basic assumption of the DMFT
procedure is therefore already incorporated. Hence, the
ansatz~(\ref{eq:ansatz}) together with the proposal to 
fit the parameters as will be described in detail below is the only
approximation necessary to obtain the $f$-electron Green function.

The coefficients $a_{\sigma}$ and $b_{\sigma}$ are determined so that
the first four ($n\in\{0,1,2,3\}$) moment sum rules
\begin{gather}
  \label{eq:moments}
  M_{\sigma}^{(n)}=
  \int \! dE \, E^n A^{(f)}_{\sigma}(E) =\langle  [
  \underbrace{[...[f_{\sigma},H]_-,...,H]_-}_{ \text{$n$-fold
      commutator}} , f_{\sigma}^{\dagger} ]_+\rangle
  \\ 
  A^{(f)}_{\sigma}(E)=-\frac{1}{\pi}\Im G^{(f)}_{ii\sigma}(E+i0^+)\nonumber
\end{gather}
are fulfilled\cite{PWN97,PHWN98}. $\Im x$ denotes the imaginary part of
$x$.

Since the moments $M_{\sigma}^{(n)}$ determine the high-energy expansion
of the Green function, the compliance of the $n=3$ sum rule
automatically leads to the correct behavior of $G_{ii\sigma}^{(f)}(E)$
up to the order $\frac{1}{E^4}$\cite{PHWN98}. Furthermore, the $n=3$ sum
rule is directly related to the correct positions and spectral weights
of the charge excitations in the strong-coupling limit
$U\rightarrow\infty$\cite{HL67,PHWN98}. This is ensured by the
occurrence of a higher-order correlation function called bandshift
$B_{\sigma}$, which is discussed in detail in the context of the Hubbard
model in reference~\onlinecite{PHWN98}, and in
reference~\onlinecite{PWN97} for the effective SIAM within the dynamical
mean-field theory.

Since we use the perturbation theory around the Hartree-Fock solution to 
determine the $\Sigma_{\sigma}^{\rm (SOC)}(E)$, another parameter enters 
the calculation, namely the chemical
potential within the Hartree-Fock calculation. 
It is
\textit{a priori} not evident that this chemical potential should be
identical to the chemical potential of the full problem. 
In fact,
several other choices to determine this parameter seem possible. Within
the iterated perturbation theory (IPT)\cite{KK96,VTJK99pre}, which is
away from half-filling also based on the ansatz (\ref{eq:ansatz}), the
Luttinger sum\cite{LW60} or equivalently the Friedel sum
rule\cite{Fri56,Lan66} is used. This, however, limits the calculation to
$T=0$ from the very beginning. We therefore define a different
constraint, demanding that the impurity occupation number within the
Hartree-Fock calculation is equivalent to the true occupation number. A
detailed investigation of this choice and other possibilities is found
in reference \onlinecite{PWN97}. Investigating the well-known
single-impurity Anderson model to test the quality of our
method\cite{MWPN99}, we found that the MPT fulfills the Friedel sum rule
within numerical accuracy not only under symmetric parameter conditions
but also in a broad range of parameters, especially when reducing the
conduction electron density. But contrary to the IPT, the MPT is
applicable also at finite temperatures.

Summarizing the features of the MPT, it should be considered
trustworthy for small $U$ since it is based on perturbation theory. But
furthermore it  is well justified in the
strong coupling regime, where the main features -- the charge excitations
are correctly reproduced. This is clearly one step beyond similar
methods which determine the parameters $a_{\sigma}$ and $b_{\sigma}$
with respect to the ``atomic'' limit
$V=0$\cite{KK96,VTJK99pre,TS99pre}.

To calculate the susceptibility, we apply an external magnetic field
$B_{\text{ext}}$ which couples to the $f$- and the $s$-electrons. The
susceptibility $\chi^{\text{(tot)}}$ is given as $\frac{\partial
M}{\partial B_{\text{ext}}}|_{B_{\text{ext}}=0}$, where $M$ is the total
magnetization of the system.  Since the $f$- and $s$-magnetization can
be computed separately, the respective contribution of the $f$-($s$-)
electrons to $\chi^{\text{(tot)}}$ can be determined.

With the above-described theory, the $f$-electron Green function and all 
quantities deriveable from it can be obtained. This includes several 
two-particle correlation functions as e.g. the $f$-electron double
occupancy $\langle
n_{\sigma}^{(f)}n_{-\sigma}^{(f)}\rangle$\cite{NolBd7,HerrmannDipl}.
However,
for the discussion below, we will also be interested in
two-particle correlation functions which are not readily obtained from
$G_{ii\sigma}^{(f)}(E)$. To determine these, as e.\ g.\ the $s$-$f$ density 
correlation function $\langle n_{\sigma}^{(f)}n_{\sigma'}^{(s)}\rangle$
we need a further approximation: We construct the following effective
medium Hamiltonian:
\begin{equation}
  \label{eq:effham}
          \begin{split}
          H^{\text{(eff)}}=&\sum_{\vec{k},\sigma} \epsilon(\vec{k})
          s_{\vec{k},\sigma}^{\dagger}s_{\vec{k},\sigma} + \sum_{i,\sigma}
          \left(\epsilon_f+\Sigma_{\sigma}(E)\right)
          f_{i,\sigma}^{\dagger}f_{i,\sigma} \\ 
          &+ V \sum_{i,\sigma}
          \left(f_{i,\sigma}^{\dagger}s_{i,\sigma}+
            s_{i,\sigma}^{\dagger}f_{i,\sigma}\right)
        \end{split}
\end{equation}
with $\Sigma_{\sigma}(E)$ being the fully self-consistent solution of
the DMFT-MPT scheme. Since the effective medium Hamiltonian is bilinear
in fermion operators, all Green functions of interest can be
evaluated exactly. By construction, the single-particle properties of
model (\ref{eq:effham}) are equivalent to the original model
(\ref{hamiltonian}) solved within the DMFT-MPT scheme. 
Although we
are aware of the fact, that using the effective Hamiltonian
(\ref{eq:effham}) must be seen as an approximation to the original model
(\ref{hamiltonian}), it is in our opinion clearly one step beyond
standard first-order perturbation theory for the latter, which would not
reproduce the non-trivial results we will discuss below. 
Let us already point out here that these non-trivial results will always 
be accompanied by special features in quantities which were derived from
the full
Hamiltonian~(\ref{hamiltonian}) using the DMFT-MPT (e.\ g.\ the
susceptibility or $\langle n_{\sigma}^{(f)}n_{-\sigma}^{(f)}\rangle$
). They seem therefore to be
not effects due to the replacement of Hamiltonian (\ref{hamiltonian})
with (\ref{eq:effham}).

\section{Results and discussion}
\label{sec:results-discussion}
\subsection{The symmetric PAM}
The symmetric PAM is defined by complete particle-hole symmetry, i.\ e.\ 
$e_f=-\frac{U}{2}$ and a particle-hole symmetric
density of states of the conduction band with the chemical potential
located at its center of
gravity.
\begin{figure}[bth]
  \begin{center}
    \epsfig{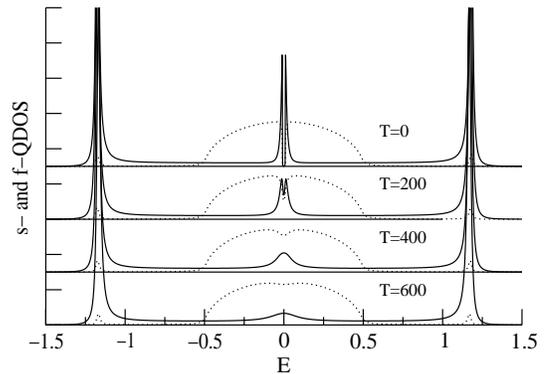}
    \caption{$s$- and $f$-densities of states for $U=2$,
      $e_f=-\frac{U}{2}=-1$, 
      $V=0.2$, $n^{(f)}=n^{(s)}=1$ and various temperatures (solid line:
      $f$-DOS, dotted line: $s$-DOS), the temperature scale is explained 
      in the text.} 
    \label{fig:qdos_sym}
  \end{center}
\end{figure}

In figure 1, the $s$- and
$f$-densities of states ($s$- and $f$-DOS) with $U=2$ and
$V=0.2$ are plotted for various temperatures. The energy scale is
defined by the free, i.e. unhybridized conduction band of unit width and
semielliptic shape centered at $E=0$. Within this energy scale, the
temperatures will be given in $\frac{K}{eV}$.
We have plotted the projections
onto the $f$($s$)-states using solid (dotted) lines.

The DOS consist of the charge excitations approximately located at
$e_f$ and $e_f+U$ which are dominantly of $f$-character and the
conduction band mostly of $s$-character which is slightly deformed due
to the hybridization.  For low temperatures, an additional feature
appears in the vicinity of $E=\mu=0$, the Kondo resonance\cite{hewson}.
It is split by the coherence gap which originates from the coherent
hybridization between $f$- and $s$-states at all lattice sites.

This coherence gap might be the theoretical
equivalent of the experimentally seen ``pseudo-gap'' e.g. in
$SmB_6$\cite{ABW79} or in the so-called Kondo insulators,
e.g. $Ce_3Bi_4Pt_3$\cite{HCTFL90} or $CeNiSe$\cite{DKJWWLF97}.
It can be understood as a level-repulsion between the conduction
band states and the effective $f$-level located at the chemical
potential. Since this effective $f$-level is clearly
correlation-induced, the coherence gap is as well.

\begin{figure}[tbh]
  \begin{center}
    \epsfig{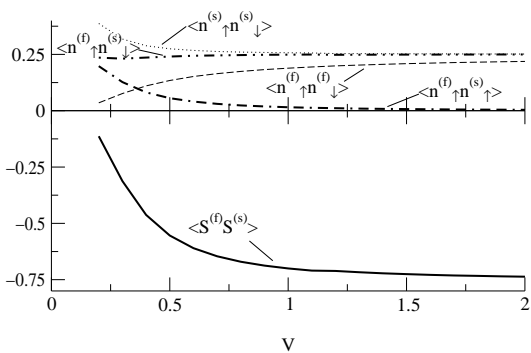}
    \caption{The on-site double-occupancy correlation functions and the
      local spin-spin correlation function $\langle
      \vec{S}_i^{(f)}\vec{S}_i^{(s)}\rangle$ as
      function of the hybridization strength $V$ for $T=0$ (other
      parameters as in fig. 1) } 
    \label{fig:v_sym}
  \end{center}
\end{figure}
The DOS obtained by our method compare very well with those calculated
by QMC in $d=1$\cite{GE98,GE99}, $d=2$\cite{GE99pre2} and
$d=\infty$\cite{Jar95,VTJK99pre}. At least for $d=1$ and $d=2$, there is
one qualitative difference, 
however. In the MPT, the Kondo resonance is of pure $f$-character,
whereas in the cited papers, the conduction band also contributes to the
resonance. Whether this is due to the maximum-entropy method\cite{JG95}
necessary to complement the QMC formalism, or an artefact of our method,
remains an open question. However, physically relevant is only the total 
density of states ($f$ plus conduction band). The choice of projecting
the DOS onto the $f$- and $s$-states, i.e. onto the basis given by the
$V=0$ solution of the problem is rather arbitrary. Therefore, the
above-discussed differences seem to be of minor importance.

Focusing on the Kondo screening problem, the
question arises on how one can observe it. One possibility found in the
literature is the 
definition of an effective magnetic moment $T\chi(T)$. This is motivated
by the Curie law which, however, only holds for high temperatures. In
the SIAM, there is a suppression of $T\chi(T)$ coinciding with the
temperature scale $T_K$ which also governs the other low-energy
properties of the system. In analogy to the SIAM, the behaviour of
$T\chi(T)$ is often interpreted as indirect indicator for the onset of
Kondo screening in the PAM\cite{Jar95,TJF97}.  Another criterion of
Kondo screening, which we will focus on, might be found in spin-spin
correlation functions. Let us look for example at the problem of two
$S=\frac{1}{2}$ spins. Of the four possible states of this system, three
are of triplet and one is of singlet nature. The spin-spin correlation
function takes the value $\langle \vec{S}_a\vec{S}_b\rangle=\frac{1}{4}$
for the triplet states and $\langle
\vec{S}_a\vec{S}_b\rangle=-\frac{3}{4}$ for the singlet state ($a$ and
$b$ denote the two spins). In the following, we will discuss only the
on-site interband spin-spin correlation function $\langle
\vec{S}_i^{(f)} \vec{S}_i^{(s)}\rangle$. It is obvious that this
function can be an indicator only for local singlet formation, as will
be discussed in more detail below.

In figure \ref{fig:v_sym}, the on-site interband spin-spin correlation
function $\langle \vec{S}_i^{(f)} \vec{S}_i^{(s)}\rangle$ as well as
several on-site double occupancy correlation functions are plotted as
function of the hybridization strength $V$ for $T=0$ (other parameters
as in figure 1).  With increasing $V$, the on-site spin-spin correlation
function approaches the value $-\frac{3}{4}$. In the same range of $V$,
the interband double occupancy with parallel spin, $\langle
n_{i\sigma}^{(f)} n_{i\sigma}^{(s)}\rangle$ vanishes, whereas the
respective correlation function with opposite spin indices, $\langle
n_{i\sigma}^{(f)} n_{i-\sigma}^{(s)}\rangle$ stays almost constant. In
analogy to the two-spin problem, these are clear indications of a local
singlet correlation.  A similar transition in the spin-spin correlation
function has been seen in a PAM with next-neighbor hybridization using
QMC, where it has been interpreted as singlet formation\cite{HMS99}.

\begin{figure}[b]
  \begin{center}
    \epsfig{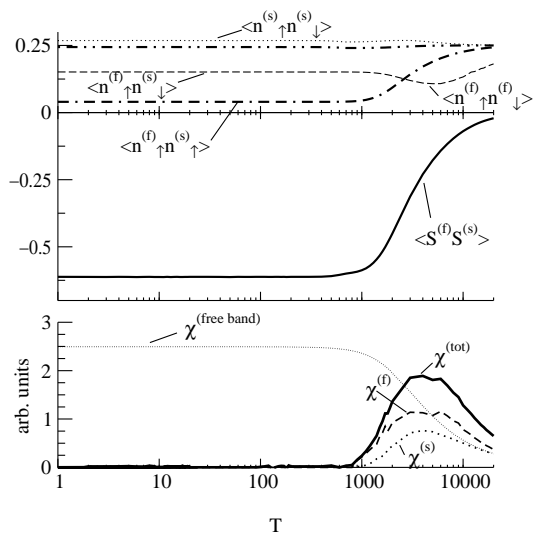}
    \caption{Upper panel: the correlation functions of figure
      \ref{fig:v_sym}, but as 
      function of temperature for $V=0.6$.
      Lower panel: the 
      susceptibility $\chi^{(tot)}$ as well as the $f$- and
      $s$-contribution to $\chi^{(tot)}$ are plotted. The thin dotted
      line ($\chi^{\text{(free band)}}$) shows the susceptibility of a
      free, i.e. unhybridized and uncorrelated conduction band with same 
      parameters as the $s$-band used in the other calculations.}
    \label{fig:t_sym_v06}
  \end{center}
\end{figure}
\begin{figure}[tbh]
  \begin{center}
    \epsfig{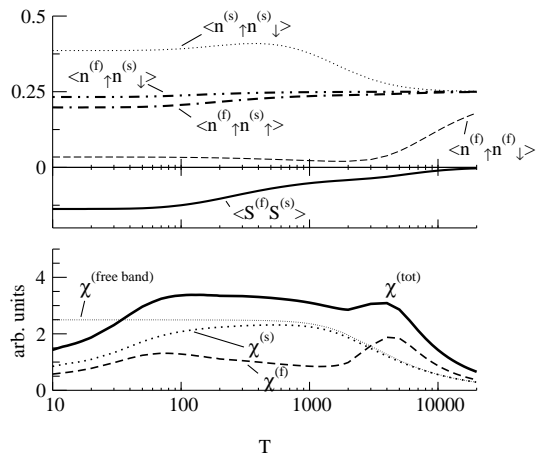}
    \caption{As figure~\ref{fig:t_sym_v06}, but with small $V=0.2$}
    \label{fig:t_sym_v02}
  \end{center}
\end{figure}

In the upper graphs of figures \ref{fig:t_sym_v06} and
\ref{fig:t_sym_v02}, the correlation functions of figure~\ref{fig:v_sym}
are plotted as function of temperature for large ($V=0.6$) and small
($V=0.2$) hybridization strengths.  Additionally, the lower graphs show
the respective susceptibilities. In the large-$V$ case, the above
described transition in the various correlation functions is clearly
visible: $\langle \vec{S}_i^{(f)} \vec{S}_i^{(s)}\rangle \rightarrow
-\frac{3}{4}$ and $\langle n_{i\sigma}^{(f)} n_{i\sigma}^{(s)}\rangle$
shows a huge drop around $T\approx 3000$, being of very small value at
low temperatures.
In the same temperature range, the susceptibility vanishes. Both
the $f$- and $s$-contribution to $\chi^{\text{(tot)}}$ disappear
simultaneously.  From these findings, we propose the occurrence of
\textit{local Kondo singlet} formation. With the term local Kondo
singlet (LKS), we want to stress, that the singlet formation is
predominantly a local process determined by the binding of one $f$- and
one $s$-electron at each lattice site. In the opposite case of
collective Kondo screening as it could occur for small $V$, the local
correlation functions discussed here need not show any
particularities. Our proposal is further supported by the behavior of
the conduction band double occupancy $\langle n_{i\sigma}^{(s)}
n_{i-\sigma}^{(s)} \rangle$. In the large-$V$ region, where we propose
the LKS formation, this correlation function is reduced compared to the
small-$V$ case, as can be clearly seen in figure \ref{fig:v_sym}. This
indicates a tendency towards localization of the conduction electrons,
which is exactly what one would expect in the case of LKS formation. The
unique temperature scale which we identify with the Kondo temperature of
$T_K\approx 1000 K$ seems to be very large. This cannot be due to the
``Hartree-Fock-character'' implied by the simple effective medium
approach of equation (\ref{eq:effham}) used to determine the higher
correlation functions, since the same temperature scale appears in the
susceptibility, which is determined from the full Hamiltonian (equation
(\ref{hamiltonian})) in the DMFT-MPT scheme. Whether $T_K$ is
over-estimated due to the mean-field character of the DMFT or the use of
the MPT, or whether it displays the true behavior of the system, would
be speculation. However, comparing our method with other means of
calculation (e.g. numerical renormalization group calculations
(NRG)\cite{BHP98,PBJ99pre}), the DMFT-MPT seems to 
overestimate energy scales. We believe this is connected to the fact that
the MPT, as any perturbative method, is unable to reproduce the
exponential temperature scale typical for the Kondo physics\cite{MWPN99}
A detailed comparison with complementary
methods, e.\ g.\ NRG, is needed to shed further light on this.

For small $V$ (figure \ref{fig:t_sym_v02}), the situation is completely
different. The systems with small and large $V$ behave similar only for
very low and very high temperatures.  For very high temperatures, a
Curie-like behavior is found as expected.  For $T=0$, the susceptibility
vanishes also in the small-$V$ case. This is compatible with the exact
result by Tsunetsugu et al.\cite{TSU97}, who proved the singlet nature
of the ground state of the symmetric PAM independent of the size of $V$
and showed the existence of a spin gap. However, for small hybridization
the various correlation functions discussed above show none of the
occurrences which led us to the conclusion of the local Kondo singlet
formation in the large-$V$ case. The susceptibility shows two
features. The one at the high temperature $T_{\text{high}}\approx 4000K$
corresponds to the delocalization of the $f$-electrons due to thermal
excitations. As can be seen at the decomposition of
$\chi^{\text{(tot)}}$, only the $f$-contribution is responsible for this
feature. The conduction electron contribution $\chi^{(s)}$ resembles at
and below $T_{\text{high}}$ still the susceptibility of a system of
uncorrelated electrons ($\chi_{\text{free band}}$). The delocalization
of the $f$-electrons is visible in the increase of $\langle
n_{i\sigma}^{(f)} n_{i-\sigma}^{(f)} \rangle$ for
$T>T_{\text{high}}$. The Kondo screening sets in at the much lower Kondo
temperature $T_K\approx 100K$. This corresponds to the temperature where
the Kondo resonance appears in the DOS (see figure
\ref{fig:qdos_sym}). As already mentioned, the local correlation
functions show here only a weak signature of Kondo screening. The Kondo
screening for small $V$ is a collective effect, the quantities
accessible by our theory do not allow more detailed investigation of
this state.

To summarize the discussion of the symmetric PAM, we find two different
kinds of Kondo singlet formation depending on the hybridization strength.
Where
for small $V$ the singlet formation involves non-local or collective
screening of the $f$-electrons by the conduction band electrons, in
the large $V$ domain, the singlet formation is dominantly a local
process. At every lattice site, one $f$- and one $s$-electron couple to
form a local Kondo singlet (LKS).

\subsection{Exhaustion problem}
One important question concerning the screening behavior is that of
exhaustion: What happens if the number of conduction electrons $n^{(s)}$ 
is reduced (the number of $f$-electrons is fixed $n^{(f)}=1$)? This
question was recently brought into  
attention by Nozi\`{e}res\cite{Noz98}. In this section, we want to present
our results concerning ``exhaustion''.

\begin{figure}[tb]
  \begin{center}
    \epsfig{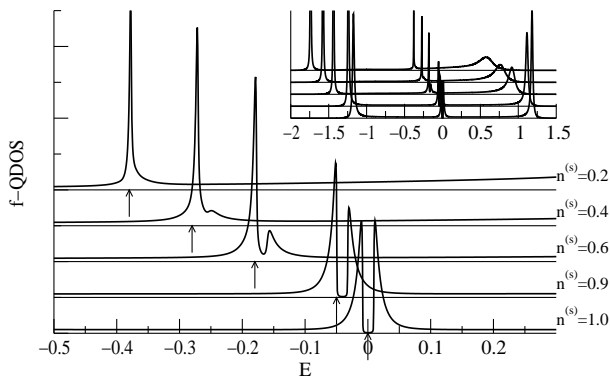}
    \caption{$f$-densities of states for $T=0$ in the vicinity of the
      chemical potential (arrows). The system
      parameters are as in figure~\ref{fig:qdos_sym} but the conduction
      band occupation is varied from $n^{(s)}=1$ to $n^{(s)}=0.2$ as
      indicated. The inset shows the $f$-DOS over the full energy range}
      \label{fig:qdos_n}
  \end{center}
\end{figure}

In figure \ref{fig:qdos_n}, the $f$-electron density of
states ($f$-DOS) is plotted for $T=0$ and various $n^{(s)}$.
The main change in the $f$-DOS is the shift of the charge excitations
as well as the Kondo resonance towards lower energies. The shift of the
charge 
excitations is due to an adjustment of $e_f$ which is necessary to keep 
the constraint $n^{(f)}=1$ with decreasing $n^{(s)}$. The Kondo
resonance also moves towards lower energies as its position is pinned to
the chemical potential $\mu$. The coherence gap stays in the vicinity 
of $\mu$, too. However, the relative position of the gap, the Kondo peak
and $\mu$ changes. Whereas for $n^{(s)}=1$, the Kondo resonance and the
gap are centered 
symmetrically around $\mu$, for $n^{(s)}<1$ the situation becomes
asymmetric. The chemical potential moves into the lower half of the
resonance. As a direct consequence, the shape of the resonance becomes
asymmetric. This strongly resembles the behavior found for the
SIAM\cite{MWPN99,hewson}. The relative shift of $\mu$ and the Kondo
resonance has also an important consequence for the coherence gap. For
$n^{(s)}<1$ the system is metallic since $\mu$ is no longer located
within the gap. With decreasing $n^{(s)}$, the gap moves further away
from the chemical potential. The imaginary part of the self-energy
$\Im\Sigma_{\sigma}(E)$ at the gap increases thereby since it shows a
``Fermi-liquid'' $E^2$ dependence around $E=\mu$.
For $n^{(s)}$ slightly below unity, the gap is still
present in form of a ``pseudo-gap''. However, for $n^{(s)}$ approaching
zero, the gap closes completely. 

This resembles the behavior
found in $Ce_3Bi_4Pt_3$ upon doping with
lanthanum\cite{HCTFL90}. The undoped system is an insulator with a very
small gap believed to be a prototypical Kondo insulator. On doping $La$, 
the system becomes a metallic heavy-fermion system. In reference
\onlinecite{HCTFL90} this is interpreted as due to the de-construction of 
lattice coherence by disorder. From the behavior of the PAM, one might also
conclude that simply the change of the electron density due to doping
would also be sufficient to explain this metal-insulator transition.  

\begin{figure}[tbh]
  \begin{center}
    \epsfig{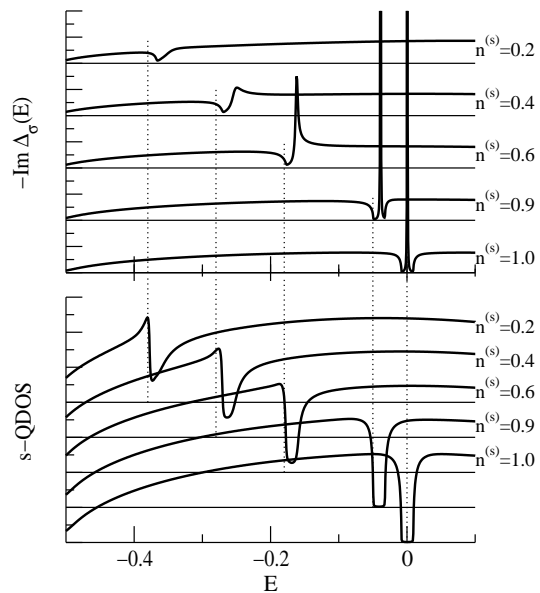}
    \caption{$s$-DOS and imaginary part of the hybridization function
      $- \Im \Delta_{\sigma} (E)$ close to the chemical potential $\mu$ for the
      same parameters as in figure~\ref{fig:qdos_n}. The dotted line
      indicates the  position of $\mu$ for the respective value of
      $n^{(s)}$. The curves for $n^{(s)}\neq 1$ have been shifted
      vertically for clarity.} 
    \label{fig:delta_n}
  \end{center}
\end{figure}
In figure~\ref{fig:qdos_n} a broadening of the upper charge excitation
is observed for $n^{(s)}\rightarrow 0.2$. Here, the upper charge
excitation overlaps with the conduction band which results in a stronger
hybridization. This could be
prevented by using a larger value for $U$. However, the impact of this
on the results discussed below is negligible.

Let us now turn to the actual problem of exhaustion. In the case of the
symmetric PAM ($n^{(s)}=n^{(f)}=1$), we have found a singlet ground
state with a finite spin gap
which is in agreement 
with QMC results\cite{BFGS87} and exact statements
concerning the ground state\cite{TSU97}. It is not clear
how the system will react when the number of
conduction electrons is reduced while  $n^{(f)}=1$ is
kept constant. One
possibility would be a partial or complete breakdown of the screening
since the conduction-band filling is not sufficient to screen
the magnetic moments of all $f$-electrons. However,
conduction-band mediated $f$-$f$ correlations could again lead to a
singlet ground state via intersite singlet correlations. This
scenario has been confirmed for the two-impurity Kondo problem\cite{ALJ95}.

Recently, the exhaustion problem has been discussed as the origin of a
new low-energy
scale\cite{Noz98,TJF97,TJF98,VTJK99pre,PBJ99pre,TJPF99}. In one of
these 
publications\cite{VTJK99pre}, the iterated perturbation theory has been
applied, which is very similar to the approximations used in this paper.
There, the authors found a gap close to the chemical potential in the
effective hybridization defined by $-\Im \Delta_{\sigma}(E)$ (see equation
(\ref{eq:seco})). In figure 
\ref{fig:delta_n}, $-\Im \Delta_{\sigma}(E)$ as well as the conduction
band density 
of states ($s$-DOS) are plotted in
the vicinity of $\mu$ for various conduction band fillings. The
parameters correspond to the DOS plotted in figure~\ref{fig:qdos_n}. In
the $s$-DOS, the 
coherence gap but no Kondo resonance is visible. The latter appears only 
in the projection onto $f$-states ($f$-DOS) and is therefore of pure
$f$-character. Again, the coherence gap shifts together with $\mu$
towards lower energies on reducing $n^{(s)}$. Let us focus on $-\Im
\Delta_{\sigma}(E)$, the upper picture in figure
\ref{fig:delta_n}. Independently
of $n^{(s)}$ there exists a gap/dip close to the chemical potential.
However, in the
symmetric case ($n^{(s)}=1$) a sharp $\delta$-like peak appears exactly
at $E=\mu$. So  
$-\Im \Delta_{\sigma}(E=\mu)$ is large in the symmetric case, but already for
any small change in $n^{(s)}$, $-\Im \Delta_{\sigma}(E=\mu)$ becomes
small since $\mu$
lies within the gap or dip. In reference~\onlinecite{VTJK99pre}, the
authors interpret   
the dip which they find for $n^{(s)}=0.4$ in the effective hybridization
as sign of the exhaustion of the conduction electrons. However, this
reduction of $-\Im \Delta_{\sigma}(E)$ around $E=\mu$, 
which we also find for $n^{(s)}=0.4$, continuously develops into a gap
for $n^{(s)}=0.9$. In the interpretation of the cited work, this would
imply that the exhaustion problem is stronger for $n^{(s)}\rightarrow 1$
than for $n^{(s)}=0.4$ since the dip evolves into a true gap.
The special case $n^{(s)}=1$ with its ``preformed gap'' surrounding the
$\delta$-like peak in $-\Im \Delta_{\sigma}(E)$ would also need
further considerations. In our opinion, without clearer justification,
this gap/dip cannot be taken as a direct sign of exhaustion.
\begin{figure}[tbh]
  \begin{center}
    \epsfig{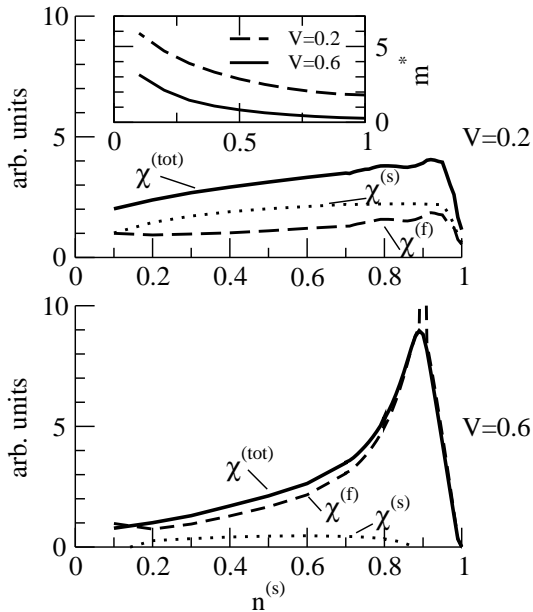}
    \caption{The $T=0$-susceptibility  and its respective $f$-level and
      conduction band contributions for small $V=0.2$ (upper panel) and
      large $V=0.6$ (lower panel) as function of conduction band
      fillig. In the inset of the upper panel, the 
      effective mass is plotted for both values of $V$. The remaining
      parameters are as in figure~\ref{fig:v_sym}.}
    \label{fig:nlauf}
  \end{center}
\end{figure}

In figure \ref{fig:nlauf}, we have plotted the zero-temperature
susceptibility for small $V=0.2$ and large $V=0.6$ as function of the
conduction band occupation number $n^{(s)}$.
In addition to that, the effective mass
$m^*=1-\frac{\partial \Sigma}{\partial E}|_{E=0}$ is plotted in the
inset of the upper panel.

\begin{figure}[tbh]
  \begin{center}
    \epsfig{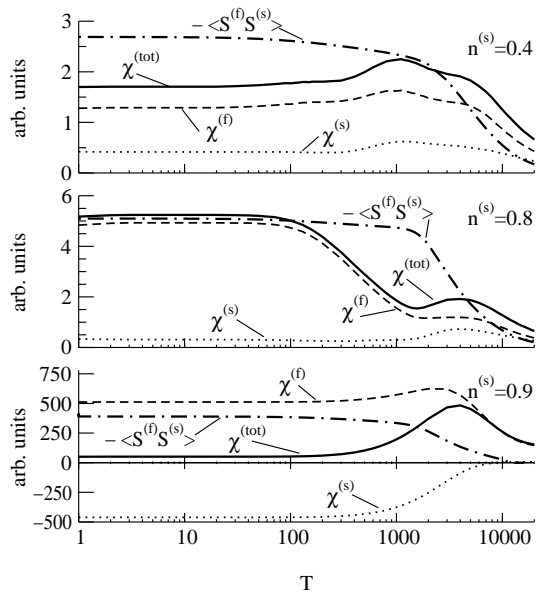}
    \caption{The susceptibility, its $f$- and $s$-contributions and
      the on-site interband spin-spin correlation function $\langle
      \vec{S}_i^{(f)}\vec{S}_i^{(s)}\rangle$ as function of temperature
      for three different conduction band occupations
      $n^{(s)}\in\{0.9,0.8,0.4\}$. The hybridization strength is
      $V=0.6$, the data corresponds to
      figure~\ref{fig:nlauf}. Please note that we have plotted the
      negative of $\langle
      \vec{S}_i^{(f)}\vec{S}_i^{(s)}\rangle$ and rescaled it by a factor 
      $10$ for the upper two, and a factor $1000$ for the lower plot.}
    \label{fig:n_t}
  \end{center}
\end{figure}
From these figures, one gains some insight in the stability of the LKS
in case of exhaustion ($n^{(s)}<n^{(f)}=1$).
In the small-$V$ case, $\chi^{(tot)}$ is roughly proportional to
the value of the DOS at $E=\mu$ for less than half-filled conduction
band. Both the $f$- and $s$-electrons contribute to
$\chi^{(tot)}$. Contrary to that, for the LKS-dominated system
($V=0.6$), only the $f$-electrons react to an external field. The
conduction band contribution $\chi^{(s)}$ is negligible for
$n^{(s)}<0.9$. So we concluse that all conduction electrons are more or
less bound into LKS states and therefore inable to react to an external
magnetic field. Another hint can be derived from the fact that $\langle
\vec{S}_i^{(f)} \vec{S}_i^{(s)}\rangle$ decreases linearly with
$n^{(s)}$ (not plotted). This can be interpreted as a linear decrease of
the number of LKS. 
I.e. all available conduction electrons form a LKS, and
$n^{(\text{bach})}=1-n^{(s)}$ $f$-electrons remain unpaired. These are
the equivalent of the ``bachelor spins'' discussed by Nozi\`{e}res for
the Kondo lattice model\cite{Noz98}. Following his argumentation, they
can be described as a system of $n^{(\text{bach})}$ spin-$\frac{1}{2}$
fermions. In the KLM these are ``hard-core''-fermions since for the
localized spins, double occupancy is strictly forbidden. In our case of
the PAM, double occupancy is principally possible, only suppressed by
the on-site repulsion $U$. Therefore, these bachelor fermions should
show similarities to a non-degenerate Hubbard model with finite $U$.
One feature of the Hubbard model is the Mott-Hubbard
transition\cite{GKKR96,gebhard}. For an exactly half-filled system with
sufficiently large $U$, the otherwise metallic system becomes insulating
due to the electron correlations. Approaching the half-filled situation
from lower carrier concentrations, the effective mass strongly
increases, and diverges at half-filling.  Bearing this in mind, the
behavior of the effective mass shown in the inset of
figure~\ref{fig:nlauf} can be understood. The most prominent feature
is the increase of $m^*$ for $n^{(s)}\rightarrow 0$. The latter,
however, implies that the above-discussed bachelor-fermion model
approaches half-filling, since the number of unpaired $f$-electrons goes
to unity. In terms of this model, the system is in the proximity of the
Mott-Hubbard transition.  Please note, that in the PAM with finite
hybridization, $n^{(s)}=0$ is only possible for extreme parameter
conditions, namely $e_f\rightarrow-\infty$ and $U\rightarrow\infty$. Our
method is not feasible for these parameters.

Let us now discuss why the susceptibility is
enlarged at $n^{(s)}=0.9$. Looking at the partial contributions
$\chi^{(f)}$ and
$\chi^{(s)}$ at $n^{(s)}=0.9$, we note that both take on large
values. However, while $\chi^{(f)}$ is positive, the conduction band
contribution becomes negative (see lowest panel of
figure~\ref{fig:n_t}). We believe this is due to the proximity of a
magnetically ordered phase. Since we do not allow for antiferromagnetic
ordering in our method, we obtain a paramagnetic system for the shown
parameters. For larger values of $U$, however, we also see an onset of
ferromagnetic order. The critical $U_c$ is lower for $n^{(s)}\approx
0.9$ than for other values of $n^{(s)}$. So the peak in the
susceptibility is in our opinion due to the proximity of the
ferromagnetic phase. The opposite signs of $\chi^{(f)}$ and $\chi^{(s)}$
result from the fact that in the ferromagnetic phase, the conduction
band will be polarized antiparallel to the $f$-levels, as will be shown
in a forthcoming publication.  However, using quantum Monte Carlo
methods, antiferromagnetic order was found for $n^{(s)}\rightarrow
1$\cite{TJF97}. In principle, the investigation of an antiferromagnetic
order is also possible within our method and is planned for future
studies. The results of reference~\onlinecite{TJF97} suggest that for
$n^{(f)}=1$ and $n^{(s)}\rightarrow 1$ the antiferromagnetic phase will
be stable against the ferromagnetic phase.

Note that also in the weakly hybridized case, this increase in $m^*$ is
visible. This is similar to the findings in
reference~\onlinecite{PBJ99pre} and can be understood in a similar
fashion as in the large-$V$ case (a detailed discussion is given in
reference~\onlinecite{TJF98}).

In figure \ref{fig:n_t}, the temperature dependence of the
susceptibility and the on-site interband spin-spin correlation function
is plotted for $V=0.6$ and $n^{(s)}\in\{0.4,0.8,0.9\}$. As can be seen,
the temperature where the LKS formation occurs, is nearly independent of
$n^{(s)}$. The LKS formation should not be confused with the potential
new low-energy scale discussed in recent
publications\cite{TJF98,PBJ99pre}. In our results, the only possible
indication for this energy scale could be the increase of
$m^*$. Complementary methods such as numerical renormalization
theory\cite{hewson,BHP98,PBJ99pre} should be used to get more insight at
this question.

\section{Conclusions}
In the present work, we have studied the periodic Anderson model (PAM)
using the modified perturbation theory in the context of the dynamical
mean-field theory. This approach is well motivated both for the large
and small coupling regime. Furthermore, applying it to the
single-impurity Anderson model\cite{MWPN99} and the Hubbard
model\cite{PWN97,WPN98,PHWN98} good accordance with (numerically) exact
methods has been found. Being fast and numerically stable, all parameter
regions of the model can conveniently be investigated, this includes
$T=0$ as well as finite temperatures.

The density of states generally consists of the charge excitations of
the localized level, the conduction band structure and, for low
temperatures, an additional peak, the Kondo resonance. At least for
symmetric parameters the latter is split by the coherence gap. The Kondo
resonance is ascribed to the phenomenon of Kondo screening meaning the
quenching of the magnetic moment of the localized levels by the
conduction electrons.

In the case of the symmetric PAM, where the $f$-level and the conduction
band are both half-filled and the two charge excitations lie
symmetrically around the chemical potential, the ground state is always
a singlet with a spin gap which is in accordance with the assumption of
complete Kondo screening. Investigating several on-site correlation
functions and the susceptibility as function of the hybridization
strength $V$, we see a crossover between two qualitatively different
regions. Whereas for small $V$, the Kondo screening is a collective
effect, for intermediate and large $V$, the screening is a dominantly
local process. At each lattice site a \textit{local Kondo singlet}
exists. It is built up by one conduction- and one $f$-electron spin.

On reducing the number of conduction electrons, the LKS remain
stable. However, due to unavailability of $s$-electrons, a finite number
of $f$-electrons are unpaired (bachelor fermions). Following a recent
reasoning by Nozi\`{e}res\cite{Noz98} and others\cite{TJF98}, the
low-temperature physics of the system should be describable by an
effective model which only regards these bachelor fermions. Our results
are compatible with this proposal.  In the susceptibility, we found
indications of the proximity of magnetically ordered phases. These will
be subject of a forthcoming paper.

\section*{Acknowledgments}
We wish to thank M.\ Potthoff for many helpful discussions.
Financial support by the Volkswagen foundation is gratefully
acknowledged. One of the authors (D.\ M.\ ) would like to thank the
Friedrich-Naumann foundation for support this work.

\end{document}